\documentclass[aps, amsfonts, amsmath,nofootinbib, showpacs, notitlepage, byrevtex, preprintnumbers, showkeys, 
prd]{revtex4-1}   
\usepackage{simplewick}
\usepackage{mathtools}
\usepackage{epsf}
\usepackage{graphicx}
\usepackage{framed}
 {\endMakeFramed}
 {\endMakeFramed}
 
\usepackage{subfigure}
\usepackage{bbm}

\newcommand{\cD}{{\cal D}}
\newcommand{\cL}{{\cal L}}

\newcommand{\cH}{{\cal H}}

\newcommand{\bea}{\begin{eqnarray}}
\newcommand{\eea}{\end{eqnarray}}
\newcommand{\be}{\begin{equation}}
\newcommand{\ee}{\end{equation}}
\newcommand{\dd}{\mathrm{d}}
\newcommand{\ii}{\mathrm{i}}
\newcommand{\efk}{\mathrm{e}}

\newcommand{\hk}{\hspace{0.1cm}}

\newcommand{\lk}{\left(}

\newcommand{\il}{\int\limits}

\newcommand{\ve}{\vec{e}}

\newcommand{\vx}{\vec{x}}

\newcommand{\vp}{\vec{p}}

\newcommand{\vA}{\vec{A}}
\newcommand{\vB}{\vec{B}}

\newcommand{\vy}{\vec{y}}
\newcommand{\vq}{\vec{q}}
\newcommand{\vb}{\vec{b}}

\newcommand{\va}{\vec{a}}

\renewcommand{\vec}[1]{\boldsymbol{#1}}

\newcommand*{\ddbar}[1][]{\mathop{\mathrm{d}\mkern-7mu\mathchar'26\mkern-1mu^{#1}}\mkern-4mu}
\newcommand*{\delbar}[1][]{\mathop{\delta\mkern-7mu\mathchar'26\mkern-1mu^{#1}}\mkern-4mu}
\newcommand{\Overset}[3][0pt]{\ensuremath{\overset{\raise#1\hbox{\scriptsize\ensuremath{#2}}}{#3}}}

\begin{document}

\title{The Wilson loop in light-front quantization}

\author{H.~Reinhardt}
\affiliation{Institut f\"ur Theoretische Physik\\
Auf der Morgenstelle 14\\
D-72076 T\"ubingen\\
Germany}
\date{\today}
%


\begin{abstract}

Using Dirac's method for the quantization of constrained systems QED is canonically quantized in the front-form in a gauge which is the light-front analog of the Weyl gauge. From the obtained vacuum wave functional the spatial Wilson loop is calculated. The result known from the canonical instant-form quantization or from the covariant path integral quantization is found only if the static limit $x^+ \to 0$ is taken in a specific order. The same ambiguity is also found in the static photon propagator in coordinate space.
\end{abstract}
\maketitle

\section{Introduction}

There are five basically different forms of realistic dynamics, which cannot be transformed into each other by a Lorentz transformation \cite{Leutwyler:1977vy}. They correspond to different subgroups of the Poincar\'e group and are defined by different parametrizations of space-time, i.e.~they differ in which coordinate is considered as ``time'' in the study of the evolution. From these five different forms so far only three have been exploited in relativistic quantum field theory \cite{R1}. These are:
i)
the instant-form, which is the usual one where time is $x^0$,
ii) the front-form, where  ``time'' is given by the light cone variable $x^+ = \frac{1}{\sqrt{2}} (x^0 + x^3)$, i.e.~surfaces of constant ``time'' are given by tangent planes to the light cone and
iii) the point-form, where ``time'' is the eigentime of a physical object, i.e.~the hypersurface of constant ``time'' are given by the hyperboloid $x^2 = \text{const} > 0$ with $x^0 > 0$.

The choice of the quantization hypersurface of constant ``time'' is restricted by microcausality in the sense that a light signal emitted from any point of the hypersurface of constant ``time'' must not cross the hypersurface. This condition is fulfilled for the instant-form and the point-form while the light-front (LF) hypersurface touches the light-cone. This is not a real problem since a signal carrying actual information moves with the group velocity which is always smaller than the phase velocity given by the speed of light. Hence, points on the light-cone cannot communicate.

According to the different parametrizations of space-time, there are different forms of Hamiltonian dynamics and thus different forms of canonical quantization. Most investigations in quantum field theory have been carried out in the instant-form quantization, which is the usual one \cite{Jackiw:1979ur}. Little work has been done so far within the point-form quantization but this method has recently been successfully used to describe hadrons within a constituent quark picture \cite*{[][{and references therein}]Melde:2007zz}. The front-form arose in the study of deep inelastic processes and has facilitated the formulation of the parton model, for a review see e.g.~ref.~\cite{Brodsky:1997de}. The front-form quantization has also proved advantageous in superstring theory \cite{Szabo:2004uy}. Recently a semi-classical approximation within the light-front quantization of QCD has been proposed, which results in a LF wave equation of hadrons which is equivalent to the equation of motion of spin modes on anti-de Sitter space \cite{Brodsky:2014yha, deTeramond:2008ht}.

The front-form dynamics has certain advantages over the instant form dynamics. The maximal number (which is seven) of generators of the Poincar\'e algebra are kinematic, i.e.~independent of the interaction, in the LF dynamics. Another simplification in the LF dynamics arises from the energy-momentum dispersion relation. For a particle with mass $m$ and $4$-momentum $p$ it is given in the instant-form dynamics  by the irrational expression
\be
\label{215-F2-x}
p_0 = \sqrt{m^2 + \vp^2} \, ,
\ee
while in the front-form dynamics it reads
\be
\label{220-f2-y}
p_+ = \frac{m^2 + \vp^2_\perp}{2 p_-} \, .
\ee
It correlates the sign of $p_+$ and $p_-$: Positivity  and finiteness of the LF energy  $p_+$ of a free massive particle requires $p_- > 0$. Zero modes $p_- = 0$ can occur only for massless particles at zero transverse momentum $\vp_\perp = 0$. Except for these zero modes the vacuum is the only state with $p_- = 0$. Since $p_-$ is kinematic, i.e.~independent
of the interaction, it is preserved at each interaction vertex and the vacuum cannot mix with the states with $p_- > 0$. As a consequence in a Feynman diagrammatic expansion of the vacuum state, loop graphs with constituents with positive $p_-$ cannot occur and the LF vacuum of the interacting theory is the trivial vacuum of the free theory. However, this argument ignores possible zero modes. 

In Yang-Mills theory, for example, the LF vacuum cannot be that of the non-interacting theory, which is the theory of free photons. This becomes evident when one considers the Wilson loop, which behaves differently in the free and in the interacting theory (perimeter vs area law). The Wilson loop is the crucial observable of the vacuum of a gauge theory since it represents an order parameter. Furthermore from this quantity the static potential between two opposite point charges can be extracted. The QCD potential is linearly rising while the potential of the free theory,  QED, is the familiar Coulomb potential.

The Wilson loop is a gauge invariant quantity, which should be the same in any gauge and in any quantization scheme. It is quite astonishing that the Wilson loop has not yet been calculated in the LF quantization. To fill this gap is the aim of the present paper. We will calculate the spatial Wilson loop in QED in LF quantization. The calculation  of the Wilson loop in LF quantization is more involved than in the usual canonical instant-form quantization, since in the former case even in QED the gauge fields at different space-time
points do not commute so that the path ordering becomes non-trivial. We will show that for the spatial Wilson loop in light-front quantization the same result as in other quantization schemes is obtained provided the static limit $x^+ \to 0$ is taken in a specific way. The same ambiguity arises also in the evaluation of the static photon propagator in coordinate space.

The organization of the rest of the paper is as follows: In the next section we redo the canonical quantization of (the photon sector of) QED in the light-front quantization in a gauge which is convenient for the evaluation of the spatial Wilson loop. In sect.~\ref{sectVC} we calculate the photon propagator, which is then used in sect.~\ref{sectVI} to calculate the spatial Wilson loop. For sake of comparison we sketch in sect.~\ref{sectIV} the evaluation of and quote the result for the Wilson loop in QED in the usual path integral and also in (instant-form) canonical quantization. Some concluding remarks are given in sect.~\ref{sectVII}. 

\section{Light-front quantization of QED}\label{sectV}

Below we carry out the quantization of QED on the light-front using Dirac's method for the quantization of constrained systems \cite{R1, Sudarshan:2016}. The quantization of QED in the LF form has been carried out before, see e.g. ref.~\cite*{[][{and references therein}]Srivastava:1999js}. Here we redo the quantization in a different gauge (see eq.~(\ref{651-19})) which is convenient for the evaluation of the Wilson loop.

The usual canonical quantization in the instant-form starts from Weyl gauge $A^0 = 0$, which fixes the gauge up to time-independent gauge transformations. The latter are then conveniently fixed by imposing the Coulomb gauge $\vec{\nabla} \cdot \vA = 0$, which is a physical gauge since the remaining transversal components of the gauge field are the gauge invariant degrees of freedom. For our purpose it turns out to be more convenient to use the light-cone analogs of the Weyl gauge and the axial gauge $A^3 = 0$.

We use the following light-cone variables
\be
\label{857-f15}
x^\pm = \frac{1}{\sqrt{2}} \lk x^0 \pm x^3 \right) \, .
\ee
In the front form $x^+$ serves as ``time'' and $x^-$ is called the longitudinal coordinate. Together with the transversal coordinates $x^{i = 1, 2}$ the light-cone variables form the 4-vector in Minkowski space $x^\mu = \{ x^+, x^1, x^2, x^- \}$ In these variables the metric tensor, defined by $(\dd s)^2 = g_{\mu \nu} \dd x^\mu \dd x^\nu$, reads
\be
\label{868-15-2}
g_{\mu \nu} = \lk \begin{array}{cccc}
0 & 0 & 0 & 1\\
0 & - 1 & 0 & 0\\
0& 0 & - 1 & 0\\
1 & 0 & 0 & 0
                  \end{array}
\right) = g^{\mu \nu}
\ee
from which follows the relations $a_\pm = a^\mp$ for any 4-vector $a^\mu$. The usual scalar product reads in the light-cone variables
\be
\label{884-15-3}
a \cdot b \equiv a^\mu b_\mu = a^+ b_+ + a^- b_- - \va_\perp \cdot \vb_\perp \, ,
\ee
where $\va_\perp = a^1 \ve_1 + a^2 \ve_2$ denotes the projection of $a^\mu$ onto the transverse plane. For the derivatives with respect to the light-cone variables $x^\pm$ we have
\be
\label{884-15-4}
\partial_\pm \equiv \frac{\partial}{\partial x^\pm} = \frac{1}{\sqrt{2}} \lk \partial_0 \pm \partial_3 \right) \, .
\ee

\subsection{Classical Hamiltonian dynamics}

We rewrite the QED Lagrangian (with $j^\mu$ being an external source)
\be
\label{844-14}
\cL = - \frac{1}{4} F_{\mu \nu} F^{\mu \nu} - A^\mu j_\mu \, , \quad \quad F_{\mu \nu} = \partial_\mu A_\nu - \partial_\nu A_\mu
\ee
in the light-cone variables $\{  x^\mu\} = \{ x^+, x^1, x^2, x^- \}$ defined by eq.~(\ref{857-f15})
\be
\label{535-1}
\cL = \frac{1}{2} F^2_{+ -} + F_{+i} F_{- i} - \frac{1}{4} F^2_{ij} - (j_+ {A^+} + j_- A^- + j_i A^i)  \, .
\ee
Here and in the following we use $i, j, k, l = 1, 2$ to label the transverse dimensions. In the Hamiltonian formulation the fields $A^\mu$ are considered as coordinates and the conjugate momenta  are defined by
\be
\label{540-2}
\Pi_\mu = \frac{\partial \cL}{\partial (\partial_+ A^\mu)} \, .
\ee
From eq.~(\ref{535-1}) we find for the momenta
\be
\label{545-3}
\Pi_+ = F_{+ -} \, , \quad \quad \Pi_- = 0 \, , \quad \quad \Pi_i = F_{i -} \, .
\ee
The classical Hamiltonian density of QED
\be
\label{550-4}
\cH = \Pi_\mu \partial_+ A^\mu - \cL
\ee
is then obtained as 
\be
\label{555-5}
\cH = \frac{1}{2} \Pi^2_+  + \frac{1}{2} B^2  - A^- \lk \partial^+ \Pi_+ + \partial^k \Pi_k - j_- \right) + j_+ A^+ + j_i A^i \, ,
\ee
where 
\be
\label{947-15}
B = \frac{1}{2} \epsilon^{ij} F_{ij}
\ee
is the component of the magnetic field perpendicular to the 1-2-plane. The momentum $\Pi_+$ can be expressed in terms of the velocity $\partial_+ A^+$ so $A^+$ is a regular dynamical coordinate. Contrary to that the momenta $\Pi_-$ and $\Pi_i$ cannot be expressed in terms of the velocities $\partial_+ A^{\mu = -, i}$, indicating that the corresponding coordinates $A^-, A^i$ are not dynamical but give rise to the primary constraints
\be
\label{567-7}
\varphi_1 : = \Pi_- \stackrel{!}{=} 0 \, \quad \quad \varphi_{2 i} : = \Pi_i - F_{i -} \stackrel{!}{=} 0 \, .
\ee
A constraint is called {\em first 
class} if it has a vanishing Poisson bracket with all constraints (including itself), otherwise it is called 
{\em second class}. 
With the usual definition of the equal-time Poisson brackets satisfying
\be
\label{576-8}
\left\{ A^\mu (\vx), \Pi_\nu (\vy) \right\} = {\delta^\mu}_\nu \delta (\vx, \vy) \, , \quad \quad \delta (\vx, \vy)
= \delta (x^- - y^-) \delta (\vx_\perp, \vy_\perp) \, 
\ee
we find from (\ref{567-7}) 
\begin{align}
\label{583-9}
\{ \varphi_1 (\vx), \varphi_1 (\vy) \} &= 0 \, , \quad \quad \{ \varphi_1 (\vx), \varphi_{2 i} (\vy) \} = 0 \, ,\nonumber\\
\{ \varphi_{2 i} (\vx), \varphi_{2 j} (\vy) \} &= - 2  \delta_{i j} \partial^x_- \delta (\vx, \vy) \, ,
\end{align}
so that $\varphi_1$ is first class while the $\varphi_{2 k}$ are second class.

The constraints (\ref{567-7}) have to be fulfilled at all times $x^+$, i.e.~they have to persist during the time evolution generated by the total Hamiltonian function
\be
\label{590-10}
H_T = \int \dd x^- \int \dd^2 x_\perp \, \cH_T (x)  
\ee
with
\be
\label{405-9}
\cH_T (x) = \cH (x) + u_m (x) \varphi_m (x) \, ,
\ee
where the $u_m (x)$ are Lagrange multiplier fields and the summation index $m$ runs over all primary constraints. This leads to the condition
\be
\label{597-11}
\partial_+ {\varphi}_m \equiv \{ \varphi_m (x), H_T \} \stackrel{!}{=} 0\, .
\ee
For a first class constraint this condition may be non-trivial and then results in a secondary constraint $\partial_+ {\varphi}_m (x) = 0$, while for a second class constraint this condition can be always fulfilled by choosing the Lagrange multiplier field $u_m (x)$ appropriately. For the second class constraints $\varphi_{2 i} (x)$ (\ref{567-7}) we find 
\be
\label{933-f16}
\partial_+ \varphi_{2 i} (x) = \left\{ \varphi_{2 i} (x), H_T \right\} = \frac{1}{2} \epsilon_{i j} \partial_j B (x) - \partial_i \Pi_+ (x) - j_i (x) + 2 \partial_- u_{2 i} (x) = 0 \, , 
\ee
which represents an equation for the Lagrange multiplier field $u_{2 i} (x)$. For the first class constraint $\varphi_1$ we find using eq.~(\ref{583-9}) 
\begin{align}
\label{604-12}
\partial_+ \varphi_1 (x) = \{ \varphi_1 (x), H_T \} &= \partial^+ \Pi_+ (x) + \partial^i \Pi_i (x)
- j_- (x) = : \varphi_3 (x) \, .
\end{align}
Since this expression is not a priori zero, it defines a new (secondary) constraint $\varphi_3 (x)$, which represents Gau\ss{}' law. One easily convinces oneself that this secondary constraint is first class
\be
\label{611-13}
\{ \varphi_1 (x), \varphi_3 (y) \}  = 0 \, , \quad \quad 
\{ \varphi_{2 i} (x), \varphi_3 (y) \} = 0 \, ,  \quad \quad 
\{ \varphi_3 (x), \varphi_3 (y) \} = 0 \, .
\ee
Actually this constraint could have already been read off from the Hamilton function (\ref{555-5}) where the expression (\ref{604-12}) is multiplied by the Lagrange multiplier field $A^- (x)$, which has no kinetic term. The first class secondary constraint $\varphi_3 (x)$ (\ref{604-12}) has to be added to the total Hamiltonian (\ref{590-10}) with an arbitrary Lagrange multiplier field $u_3 (x)$ resulting in the extended Hamiltonian
\be
\label{987-x1}
H_E = H_T + \int \dd^3 x \, u_3 (x) \varphi_3 (x) \equiv \int \dd^3 x \, \cH_E \, .
\ee
The Gau{\ss} law term $(-A^- \varphi_3)$ of the original Hamiltonian function (\ref{555-5}) can be absorbed into the second term of $H_E$ (\ref{987-x1}) by redefining  the Lagrange multiplier field
\be
\label{994-f16a}
u_3 (x) \to u'_3 (x) = u_3 (x) - A^- (x) \, .
\ee
The extended Hamiltonian density (\ref{987-x1}) then reads
\begin{align}
\label{999-x2}
\cH_E = &  \frac{1}{2} \Pi^2_+ + \frac{1}{2} B^2 + u_1 (\vx) \varphi_1 (\vx) + u_{2 i} (\vx) \varphi_{2 i} (\vx) + u'_3 (x) \varphi_3(x) + j_+ A^+ + j_i A^i \, .
\end{align}
It is easy to check the persistency of the constraint $\varphi_3 (x)$ (\ref{604-12}). Using eq.~(\ref{611-13}) we find after a short calculation
\be
\label{641-18}
\partial_+ \varphi_3 (x) = \{ \varphi_3 (x), H_E \} = \{ \varphi_3 (x), H \} = 0  \, ,
\ee
so that the time-evolution of $\varphi_3 (x)$ 
does not give rise to a new constraint. Thus altogether we have the constraints $\varphi_1, \varphi_{2 i}$ (\ref{567-7}) and $\varphi_3$ (\ref{604-12}), where $\varphi_1$ and $\varphi_3$ are first class and $\varphi_{2 i}$ are second class.

\subsection{Gauge fixing and resolution of the constraints}

First class constraints represent weak operator identities which have to be imposed as subsidiary condition to the physical states. It is convenient to degrade all first class constraints to second class ones. This is achieved by imposing appropriate gauge conditions. In the present case two convenient gauge conditions are
\be
\label{651-19}
\chi_1 : = A^- = 0  \, , \quad \quad \chi_2 : =  A^+ = 0\, .
\ee
The first gauge condition, $A^- = A_+ = 0$, is the light-front analog of Weyl gauge and fixes the gauge up to $x^+$-independent gauge transformations. The latter are fixed by the condition $A^+ = A_- = 0$, which is the light-front analog of the axial gauge. The two gauge conditions (\ref{651-19}) turn all first class constraints into second class ones as we will show below. For this purpose it is convenient to collect all constraints and gauge conditions into a single vector
\be
\label{668-21}
\phi_1 = \varphi_1 \, , \quad \quad \phi_{2 i} = \varphi_{2 i} \, \quad \quad \phi_3 = \varphi_3 \, , \quad \quad \phi_4 = \chi_1 \, , \quad \quad \phi_5 = \chi_2 \, .
\ee
Using eq.~(\ref{576-8}) straightforward calculations yield for the matrix 
\be
\label{674-22}
C_{mn} (x, y) = \{ \phi_m (x), \phi_n (y) \} \, .
\ee
the explicit expression
\be
\label{679-23}
C_{i j} (\vx, \vy) = \lk \begin{array}{ccccc}
                         0 & 0 & 0 & - \delta (x, y) & 0 \\
                         0 & - 2 \delta_{i j} \partial^x_- \delta (x, y)  & 0 & 0 & 0 \\
                         0 & 0 & 0 & 0 & - \partial^x_- \delta (x, y) \\
                         \delta (x, y) & 0 & 0 & 0 & 0 \\
                         0 & 0 & - \partial^x_- \delta (x, y) & 0 & 0 \\
                                                 \end{array}
                                                 \right) \, .
\ee
This matrix is non-singular and hence all constraints including the gauge conditions are now indeed second class. Inverting this matrix yields
\be
\label{682-24}
C^{- 1}_{i j} (x, y) = \lk \begin{array}{ccccc}
                         0 & 0 & 0 & \delta (x, y) & 0 \\
                         0 & \frac{1}{2}  \delta_{i j} \partial^x_- G (x, y)  & 0 & 0 & 0 \\
                         0 & 0 & 0 & 0 & \partial^x_- G (x, y) \\
                         - \delta (x, y) & 0 & 0 & 0 & 0 \\
                         0 & 0 & \partial^x_- G (x, y) & 0 & 0 \\
                                                 \end{array}
                                                 \right) \, 
\ee
Here $G (x, y)$ is the Green's function of the operator  $- \partial^2_-$, for which one finds
\be
\label{693-25}
G (x, y) : = - \frac{1}{\partial^2_-} \delta (x, y) = - \frac{1}{2} | x^- - y^- | \delta
(\vx_\perp, \vy_\perp)
\ee
Furthermore, in eq.~(\ref{682-24}) we have used the fact that the inverse of the kernel $K (x, y) = - \partial^x_- \delta (x, y)$ is given by $K^{- 1} (x, y) = \partial^x_- G (x, y)$, for which we find from (\ref{693-25})
\be
\label{699-26}
\partial_- G (x, y) = - \frac{1}{2} \text{sign} (x^- - y^-) \delta (\vx_\perp, \vy_\perp) \, .
\ee
With the inverse matrix (\ref{682-24}) at hand we can now evaluate the {\em Dirac brackets}
\be
\label{704-27}
\{ A, B \}_D = \{ A, B \} - \int \dd^3 x \int \dd^3 y \, \{ A, \phi_m (x) \} C^{- 1}_{mn} (x, y) \{ \phi_n (y), B \} \, .
\ee
The relevant Poisson brackets of the constraints with the canonical variables are listed in Appendix \ref{appC}. Using these expressions one finds then the following Dirac brackets of the canonical variables 
\begin{align}
\label{710-28}
\{ A^i (x), A^j (y) \}_D & = \frac{1}{2} \delta^{i j} \partial^x_- G (x, y)  \\
\label{713-29}
\{ A^i (x), \Pi_j (y) \}_D &= \frac{1}{2} \delta^i_j \delta (x, y)\\
\label{715-30}
\{ \Pi_i (x), \Pi_j (y) \}_D &= \frac{1}{2} \delta_{i j} \partial^x_- \delta (x, y) \, , \\
\label{782-x13}
\{ A^i (x), \Pi_+ (y) \}_D &= - \frac{1}{2} \partial^x_i \partial^x_- G (x, y) \, .
\end{align}
The characteristic feature of the light-front dynamics is that neither the fields nor the momenta have vanishing Dirac brackets with themselves. 

Once the Dirac-brackets have been evaluated all constraints (and gauge conditions) have to be imposed. With $\varphi_m(x) = 0$ the total Hamiltonian function $\cH_E$ (\ref{999-x2}) reduces to
\be\label{davide}
\cH_E = \frac{1}{2} \Pi_+^2 + \frac{1}{2} B^2 + j_+ A^+ + j_i A^i \, .
\ee
In deriving the expression (\ref{555-5}) for $\cH$ the constrains $\varphi_1, \varphi_{2 i}$ (\ref{567-7}) have already been imposed. We still have to impose the constraint $\varphi_3 (x) = 0$ (\ref{604-12}), i.e.~Gau\ss{}' law
\be
\label{821-f14}
\partial_- \Pi_+ - \partial_i \Pi_i - j_- = 0 \, , 
\ee
which can be solved for $\Pi_+$ to yield
\be
\label{826-x14-1}
\Pi_+ = \frac{1}{\partial_-} (\partial_i \Pi_i - j_-) \, .
\ee
It is interesting to note that this expression for $\Pi_+$ preserves the Dirac bracket (\ref{782-x13}): Inserting the r.h.s.~of eq.~(\ref{826-x14-1}) for $\Pi_+$ into (\ref{782-x13}) and using eq.~(\ref{713-29}) the r.h.s.~of eq.~(\ref{782-x13}) is recovered.

Inserting the expression (\ref{826-x14-1}) for $\Pi_+$ into the Hamiltonian density $\cH_E$, eq.~(\ref{davide}), we obtain the classical Hamiltonian density of electrodynamics in the front-form in the gauge (\ref{651-19}) 
\be
\label{838-xxx2}
\cH_E = \frac{1}{2} (\partial_i \Pi_i - j_-) (- \partial^2_-)^{- 1} (\partial_j \Pi_j - j_-) + \frac{1}{2} B^2 + j_+ A^+ + j_i A^i \, .
\ee
At first sight this Hamiltonian density looks simpler than its counterpart in the instant form since only two field degrees of freedom $A_{i = 1, 2}$ and their momenta explicitly enter. Furthermore, the potential term is given by the third component of the magnetic field $B = \partial_1 A_2 - \partial_2 A_1$ alone. We should also note that while only the transverse part of the gauge potential enters the potential the kinetic term contains only the longitudinal part of the momentum operator. This is quite strange and can yield a non-trivial 
dynamics only because of the unusual Dirac brackets (\ref{710-28}) $\ldots$ (\ref{782-x13}). 

\subsection{The quantum theory}

The quantization proceeds by promoting the canonical variables $A, \Pi$ to operators $\hat{A}, \hat{\Pi}$ and imposing commutation relations given by $\ii$ times the corresponding Dirac-brackets
\be
\label{722-31}
[\ldots , \ldots] \stackrel{!}{=} \ii \{ \ldots , \ldots \}_D \, .
\ee
Therefore, the above obtained Dirac brackets (\ref{710-28}, \ref{715-30}) imply that in the quantum theory neither the fields nor the momenta commute among themselves:
\begin{align}
 \label{768-f13}
 \left[ A^k (\vx), A^l (\vy) \right] &= - \frac{\ii}{4} \delta^{kl} \text{sign} (x^- - y^-) \delta^2 (\vx_\perp - \vy_\perp) \nonumber\\
 \left[\Pi_k (\vx), \Pi_l (\vy) \right] &= \frac{\ii}{2} \delta_{kl} \partial^x_- \delta (x^- - y^-) \delta^2 (\vx_\perp - \vy_\perp) \, .
\end{align}
Also the commutation relation between the fields and momenta
\be
\label{7310-32}
[A^k(\vx), \Pi_l (\vy) ] = \ii \frac{1}{2} \delta^k_l  \delta (x^- - y^-)  \delta^2 (\vx_\perp - \vy_\perp) \, ,
\ee
differ by a factor of $1/2$ from the usual one. All three commutation relations can be realized by the following decomposition of fields and momenta in terms of creation and annihilation operators
\begin{align}
\label{814-x1}
A^k (\vx) &= \int \ddbar^3 p \frac{\Theta (p_-)}{\sqrt{2 p_-}} \left[ a_k (\vp) \efk^{\ii \vp \cdot \vx} + a^\dagger_k (\vp) \efk^{- \ii \vp \cdot \vx} \right] \, , \\
\label{817-x2}
\Pi_k (\vx) &= - \ii \int \ddbar^3 p \, \Theta (p_-) \sqrt{\frac{p_-}{2}} \left[a_k (\vp) \efk^{\ii \vp \cdot \vx} - a^\dagger_k(\vp) \efk^{- \ii \vp \cdot \vx} \right] \, 
\end{align}
provided the latter satisfy the usual Bose commutation relations with the non-vanishing commutator given by
\begin{align}
\label{833-f15-2-x3}
[a_k (\vp), a^\dagger_l (\vq)] &= \delta_{kl} \delbar^3 (\vp - \vq) \, .
\end{align}
Here we use the notation $\vx = (\vx_\perp, x^-)$ and $\vp = (\vp_\perp, -p_-)$ so that
\be
\label{1301-23}
\vx \cdot \vp = \vx_\perp \cdot \vp_\perp - x^- p_- \, .
\ee
Furthermore we have defined
\be
\label{1418-5}
\int \ddbar^3 p = \int \ddbar   p_- \int \ddbar^2 p_\perp \, , \qquad \ddbar = \frac{\dd}{2 \pi}
\ee
and
\be
\label{1311-23-1}
\delbar^3 (\vp) = (2 \pi)^3 \delta^3 (\vp) =  (2 \pi)^3 \delta (p_-) \delta^2 (\vp_\perp) \, .
\ee
With the decompositions (\ref{814-x1}), (\ref{817-x2}) one finds from eq.~(\ref{838-xxx2}) for the QED Hamiltonian in LF quantization the Fock-space representation
\be
\label{1033-f15-7}
H = \delbar^3 (0) \il^\infty_0 \ddbar p_- \int \ddbar^2 p_\perp \, \omega (\vp) + \il^\infty_0 \ddbar p_- \int \ddbar^2 p_\perp \, \omega (\vp) a^\dagger_k (\vp) a_k (\vp) \, ,
\ee
where
\be
\label{1018-f15-4}
\omega ({\vp}) = \frac{\vp^2_\perp}{2 p_-}
\ee
is the photon energy in the front-form and we have put the external currents to zero $j^\mu = 0$, since they are not needed in the following. This is the usual second quantized form of the QED Hamiltonian known from instant-form quantization except that the true photon energy $\omega (\vp) = | \vp|$ is replaced here by its front-form (\ref{1018-f15-4}). The ground state of this Hamiltonian is obviously given by the Fock space vacuum $| 0 \rangle$, which is defined by $a_k (\vp) | 0 \rangle = 0$. The coordinate representation $\psi (A) = \langle A | 0 \rangle$ of this state is not available since we do not have the coordinate representation of the momentum operator $\vec{\Pi}$ at our disposal. However, as in the case of the harmonic oscillator, we can express all observables via eqs.~(\ref{814-x1}), (\ref{817-x2}) in terms of the creation and annihilation operators and work exclusively in Fock-space. For the static propagator
\be
\label{744-16c}
\cD^{kl}_{BC} (\vx - \vy) = \langle 0 | B^k (\vx) C^l (\vy) | 0 \rangle = \int \ddbar^3 p \, \efk^{\ii \vp \cdot (\vx - \vy)} \cD^{kl}_{BC} (\vp)
\ee
of the canonical variables $B^k, C^k \in \{ A^k, \Pi^k\}$ we find in momentum space
\begin{align}
\label{1149-20}
\cD^{kl}_{AA} (\vp) &= \delta^{kl} \frac{\Theta (p_-)}{2 p_-} \, ,
\end{align}
\begin{align}
\label{1000-f18-2-3}
\cD_{\Pi \Pi}^{kl} (\vp) &= \frac{1}{2} \delta^{kl} p_- \Theta (p_-) \, , \quad \quad \cD_{A \Pi}^{kl} (\vp) = \frac{1}{2} \ii \Theta (p_-) \delta^{kl} \, .
\end{align}
The above result (\ref{1149-20}) should be compared with the static photon propagator in the instant-form quantization, e.g.~in Coulomb gauge (see sect.~\ref{sectIV})
\be
\label{994-f18-3}
\cD^{kl}_{AA} (\vp) = t^{kl} (\vp)  \frac{1}{2 \omega (p)} \, , \qquad \cD^{kl}_{\Pi \Pi} (\vp) = \frac{1}{2} t^{kl} (\vp) \omega (p) \, , \qquad \cD^{kl}_{A \Pi} (\vp) =  \ii \frac{1}{2} t^{kl} (\vp) \, ,
\ee
where $t^{kl} (\vp)$ is the transversal projector and  $\omega (p) = | \vp |$ represents the true photon energy. In the light-front quantization (\ref{1149-20}), (\ref{1000-f18-2-3}) the photon energy, $| \vp |$, is replaced by $p_-$ and the propagators exist only for $p_- > 0$.  

\section{The photon propagator}\label{sectVC}

The canonical quantization carried out above provides us with the time-independent wave functionals arising from the solution of the stationary Schr\"odinger equation. With the stationary  wave functionals at our disposal time dependent processes are conveniently described in the Heisenberg picture. This is, in particular, true for the photon propagator defined by
\be
\label{1457-f26}
\cD^{\mu \nu} (x, y)  = \langle 0 | T_+ A^\mu (x) A^\nu (y) | 0 \rangle \, ,
\ee
where $T_+$ denotes time-ordering with respect to the light-front time $x^+$, $| 0 \rangle$ is the (stationary) vacuum wave functional and 
\be
\label{1393-x0}
A^k (x) \equiv A^k (x^+, \vx) = \efk^{\ii H x^+} A^k (\vx) \efk^{- \ii H x^+} \, ,
\ee
is the field operator in the Heisenberg picture. Here $H$ is the LF Hamiltonian (\ref{1033-f15-7}). With the commutation relation (\ref{833-f15-2-x3}) and the explicit form of the QED Hamiltonian  (\ref{1033-f15-7}) we find for the time-dependent creation and annihilation operators
\begin{align}
\label{1399-x1}
a_k (x^+, \vp) &: = \efk^{\ii H x^+} a_k (\vp) \efk^{- \ii H x^+} = a_k (\vp) \efk^{- \ii \omega (\vp) x^+} \\
\label{1401-x2}
a^\dagger_k (x^+, \vp) & := \efk^{\ii Hx^+} a^\dagger_k (\vp) \efk^{- \ii H x^+} = \efk^{\ii \omega (\vp) x^+} a^\dagger_k (\vp) \, ,
\end{align}
where $\omega (\vp)$ is the LF photon energy (\ref{1018-f15-4}). From eq.~(\ref{833-f15-2-x3}) the commutation relation
\be
\label{1500-f27}
\left[ a_k (x^+, \vp), a_l (y^+, \vq) \right] = \delta_{kl} \delbar_{kl} (\vp- \vq) \efk^{- \ii \omega (p) (x^+ - y^+)}
\ee
follows. With eqs.~(\ref{1399-x1}) and (\ref{1401-x2}) the time dependent photon field (\ref{1393-x0}) is obtained from eq.~(\ref{814-x1}) as
\begin{align}
\label{1412-4}
A^k (x) & = \int \ddbar^3 p \frac{\Theta (p_-)}{\sqrt{2 p_-}} \left[ a_k (x^+, \vp) \efk^{\ii \vp \cdot \vx} 
+ a_k^\dagger (x^+, \vp) \efk^{- \ii \vp \cdot \vx} \right] \, 
\end{align}
and an analogous expression is found from eq.~(\ref{817-x2}) for the momentum operator. From this representation with $a_k (\vp) | 0 \rangle = 0$ we find for the photon two-point function
\be
\label{1433-8}
\langle 0 | A^k (x) A^l (y) | 0 \rangle
= \delta^{kl} \int \ddbar^3 p \frac{\Theta (p_-)}{2 p_-} \efk^{- \ii \omega (\vp) (x^+ - y^+)} 
\efk^{\ii \vp \cdot (\vx - \vy)}  \, .
\ee
Note that this amplitude is restricted to the spatial components $k, l = 1, 2$ due to the use of the gauge $A^- = 0 = A^+$. With the explicit form of the photon energy (\ref{1018-f15-4}) the integral over $\vp_\perp$ is a usual Fresnel-integral
\be
\label{1443-9}
\int \ddbar^2 p_\perp \, \efk^{- \ii \frac{x^+}{2 p_-} \vp^2_\perp + i \vx_\perp \cdot \vp_\perp} = \lk \sqrt{\frac{p_-}{2 \pi \ii x^+}} \right)^2 \efk^{\ii \frac{\vx^2_\perp}{2 x^+} p_-} \, .
\ee
We are then left with the following representation of the photon two-point function in LF quantization
\be
\label{1449-10}
\langle 0 | A^k (x) A^l (y) | 0 \rangle =  \frac{1}{2} \delta^{kl} \frac{1}{4 \pi^2 \ii (x^+ - y^+)} \il^\infty_0 \dd p_- \, \exp\left[-\ii  p_- \lk x^- - y^- - \frac{(\vx_\perp - \vy_\perp)^2}{2 (x^+ - y^+)}\right)\right] \, .
\ee
The remaining integral is well defined only after introducing a damping term, replacing  $x^- \to x^- - \ii \varepsilon \, , \varepsilon \to 0$. This yields
\be
\label{1455-11}
\langle 0 | A^k (x) A^l (y) | 0 \rangle = - \delta_{k l} \frac{1}{4 \pi^2 \left[(x  - y)^2 - \ii \varepsilon \, \text{sign} \, (x^+ -  y^+)\right]} \, ,
\ee
where $x^2 = 2 x_+ x_- - \vx^2_\perp$. From this expression we find for the photon propagator (\ref{1457-f26})
\begin{align}
\label{1428-7}
\cD^{kl} (x - y) & = \Theta (x^+ - y^+) \langle 0 | A^k (x) A^l (y) | 0 \rangle + \Theta (y^+ - x^+) \langle 0 | A^l (y) A^k (x) | 0 \rangle
\end{align}
the compact form
\be
\label{1465-13}
\cD^{kl} (x) = \frac{-\delta^{kl} }{4 \pi^2 x^2 - \ii \varepsilon} \, .
\ee
This is precisely the usual photon propagator obtained in Landau gauge within the functional integral quantization (see eq.~(\ref{2473-f45-x5}) below) except that the transversal projector is replaced here by the Kronecker symbol.

From eq.~(\ref{1465-13}) we find for the ``static'' photon propagator in LF quantization
\be
\label{1477-x1}
\left. \cD^{ij} (x) \right|_{x^+ = 0} = \frac{\delta^{ij}}{4 \pi^2 \vx^2_\perp + \ii \varepsilon} \, .
\ee
Although we have merely taken the limit $x^+ \to 0$ this propagator is also independent of $x^-$. Except for the absence of the transverse projector in eq.~(\ref{1477-x1}) the same static photon  propagator is obtained in the usual canonical instant-form quantization in Coulomb gauge if one puts $x^3 = 0$ in addition to $x^0 = 0$, see eq.~(\ref{2550-f49-x14}) below. The fact that the static LF propagator is independent of $x^-$ should simplify the LF description. 

One important note is in order: To arrive at the static LF photon propagator it was crucial to start with the time-dependent fields and take the limit $x^+ \to 0$ only after the momentum integrals were carried out. To see this let us alternatively try to evaluate the static propagator by putting $x^+ = y^+$ immediately in eq.~(\ref{1433-8}) resulting in 
\begin{align}
\label{1483-26}
\cD^{kl} (x^+ = 0, \vx) &= \delta^{kl} \il^\infty_0 \frac{\dd p_-}{2 p_-} \int \ddbar^2 p_\perp \, \efk^{- \ii p_- x^-} \efk^{\ii \vp_\perp \cdot \vx_\perp} = \delta^{kl} \delta^2 (\vx_\perp) \il^\infty_0 \frac{\dd p_-}{2 p_-} \, \efk^{- \ii p_- x^-}  \, .
\end{align}
Contrary to eq.~(\ref{1477-x1}) this result depends on $x^-$ and is ill-defined. The lesson from this calculation is that the limit $x^+ \to 0$ can be taken only after the momentum integrals $\int \ddbar p_\perp$ and $\int \dd p_-$ have been carried out. Keeping $x^+ \neq 0$ during the calculation even for observables which are $x^+$-independent obviously represents some regularization in the spirit of the point splitting. This will be important later in the evaluation of the Wilson loop to be given in section \ref{sectVI}.

For later application let us also calculate the commutator for the time-dependent gauge fields. Since the gauge field (\ref{1412-4}) is linear in the creation and annihilation   operators the commutator $[A, A]$ is a $c$-number. Therefore we have with $\langle 0 \vert 0 \rangle = 1$
\begin{align}
\label{1551-f28}
\left[A^k (x), A^l (x) \right] &= \langle 0 | \left[ A^k (x), A^l  (y) \right] | 0 \rangle \nonumber\\
&= \langle 0 | A^k (x) A^l (y) | 0 \rangle - \langle 0 | A^l (y) A^k (x) | 0 \rangle \, .
\end{align}
From eq.~(\ref{1455-11}) we find
\be
\label{1557-f28-1}
[A^k (x), A^l (y) ] = - \ii \delta^{kl} \frac{1}{2 \pi^2} \frac{\varepsilon \text{sign} (x^+ - y^+)}{[(x - y)^2]^2 + \varepsilon^2} \, .
\ee
Using
\be
\label{1562-f28-2}
\lim\limits_{\varepsilon \to 0} \frac{\varepsilon}{x^2 + \varepsilon^2} = \pi \delta (x)
\ee
we obtain
\be
\label{1538-f27-6}
[A^k (x), A^l  (y)] = - \frac{\ii}{2 \pi} \delta^{kl} \text{sign} (x^+ - y^+) \delta \lk (x - y)^2 \right) \, .
\ee
In the limit $x^+ \to y^+$ this expression should reduce to the equal-time commutation relation (\ref{768-f13}). This is, however, not obvious, since we cannot take this limit straightforwardly due to sign function. Furthermore the $\delta$-functions are not the same in both expressions. However, the $\delta$-function in (\ref{1538-f27-6}) is non-zero only for $2 (x^+ - y^+) (x^- - y^-) - \vx^2_\perp = 0$, which requires that the sign of $(x^- - y^-)$ is the same as that of $(x^+ - y^+)$. Therefore  in eq.~(\ref{1538-f27-6}) we can replace $\text{sign} (x^+ - y^+)$ by $\text{sign} (x^- - y^-)$. Then the equal time limit $x^+ \to y^+$ can be taken yielding
\be
\label{1551-f27-y1}
[A^k (x), A^l (y)]_{x^+ = y^+} = - \frac{\ii}{2 \pi} \delta^{kl} \text{sign} (x^- - y^-) \delta \lk (\vx_\perp - \vy_\perp)^2 \right) \, .
\ee
This is not yet the expression found in eq.~(\ref{768-f13}). To see the equivalence between eqs.~(\ref{768-f13}) and (\ref{1551-f27-y1}) let us set $\vec{r}_\perp = \vx_\perp - \vy_\perp$ and use polar coordinates $\vec{r}_\perp (r, \varphi)$. The $\delta$-function $\delta^2 (\vec{r}_\perp)$ in eq.~(\ref{768-f13}) is non-zero only if the module $r = | \vec{r}_\perp|$ vanishes. However, for $r = 0$ the angle $\varphi$ is indefinite so that the following identity holds
\be
\label{1559-f27a-7}
\delta^2 (\vec{r}_\perp) = \frac{1}{r} \delta (r) \frac{1}{\pi} \, .
\ee
Indeed, both sides are non-zero (and also divergent) only for $r = 0$ and have the same normalization
\be
\label{1564-f27a-8}
\int \dd^2 r_\perp \, \delta^2 (\vec{r}_\perp) = \il^\infty_0 \dd r r \il^{2 \pi}_0 \dd \varphi \, \frac{1}{r} \delta (r) \frac{1}{\pi} = 2 \il^\infty_0 \dd r \, \delta (r) = 1 \, .
\ee
With $\delta (\vec{r}^2_\perp) \equiv \delta (r^2) = \frac{1}{2 r} \delta (r)$ we eventually find from (\ref{1559-f27a-7})
\be
\label{1570-f27-9}
\delta (\vec{r}^2_\perp) = \frac{\pi}{2} \delta^2 (\vec{r}_\perp) \, ,
\ee
which establishes the equivalence between eq.~(\ref{1551-f27-y1}) and (\ref{768-f13}). 

\section{The Wilson loop in QED}\label{sectIV}

The (Wegner-)Wilson loop is defined as the path ordered exponent
\be
\label{2254-41x1}
W [A] = P \exp \left[ - \ii g \oint\limits_C \dd x^\mu A_\mu (x) \right] \, ,
\ee
where $g$ is the electric charge and $C$ is a closed loop in Minkowski space. Its vacuum expectation value $\langle W [A] \rangle$ serves as order parameter in gauge theory and allows to extract the potential between two opposite static point charges. In QED, in the usual path integral quantization and also in the canonical quantization in the instant-form, the gauge fields at different space-time points commute and the path ordering becomes irrelevant. This, however, is not the case in the front-form quantizations, see eq.~(\ref{1551-f27-y1}). 

\subsection{Functional integral approach}

In the usual path integral formulation of QED the vacuum expectation values are defined by
\be
\label{2448-f45-x1}
\langle \cdots \rangle = \int \cD A_\mu (x) \ldots \efk^{\ii S [A]} \, ,
\ee
where $S [A] = \int \dd^4 x \, \cL (x)$
is the classical action and $\cL$ is defined in eq.~(\ref{844-14}). The functional integral in eq.~(\ref{2448-f45-x1}) is well defined only after gauge fixing, which is usually done by means of the Faddeev-Popov method. Using the familiar Landau (Lorentz) gauge $\partial_\mu A^\mu = 0$ the Faddeev-Popov determinant $\text{Det} (- \Box)$ is an irrelevant constant which can be absorbed into the normalization of the functional integral, which extends over the transversal components of the gauge field only. Since the classical action $S [A]$ is quadratic in the gauge field  the Wilson loop can be evaluated in closed form yielding
\be
\label{2463-f45-x3}
\langle W \rangle = \exp \left[ - \frac{g^2}{2} \oint\limits_C \dd x^\mu \oint\limits_C \dd y^\nu \, \cD_{\mu \nu}(x - y) \right] \, .
\ee
Here
\be
\label{2468-f45-x4}
\cD_{\mu \nu} (x - y) = \langle A_\mu (x) A_\nu (y) \rangle
\ee
is the photon propagator, for which one obtains in Landau gauge $\partial_\mu A^\mu = 0$
\be
\label{2473-f45-x5}
\cD_{\mu \nu} (x) = t_{\mu \nu} (x) \frac{-1}{4 \pi x^2 - \ii \varepsilon} \, ,
\ee
where $t_{\mu \nu} = - g_{\mu \nu} - \partial_\mu \partial_\nu / \partial^2$ is the transverse projector. Only the metric tensor $g_{\mu \nu}$ part of this projector contributes in the exponent of the Wilson loop (\ref{2463-f45-x3}).

A temporal loop describes the creation of a charge-anticharge pair which evolves in time and is eventually annihilated. Using a rectangular temporal loop with length $T$ along the time axis and with spatial extension  $R$ one extracts from $\ln \langle W \rangle$ in the limit $T \gg R$ the familiar Coulomb law \cite{Becher:1984my}
\be
\label{2482-f45-x6}
\lim\limits_{T \to \infty} \ln \langle W \rangle / \ii T = - \frac{g^2}{4 \pi R} + \mbox{divergent self-energy terms independent of} \, R .
\ee
The same result is also found in the usual canonical quantization in the instant-form given below.

\subsection{Canonical quantization in the instant-form}\label{sectIVB}

The usual canonical quantization in the instant-form is carried out in Weyl gauge $A^0 = 0$ resulting in the Hamiltonian
\be
\label{2495-f49-x7}
H = \frac{1}{2} \int \dd^3 x \lk \vec{\Pi}^2 (x) + \vB^2 (x) \right) \, ,
\ee
where
$
\vec{\Pi} (x) = \delta / \ii \delta \vA (x)
$
is the momentum operator, which represents the electric field, and $\vB = \vec{\nabla} \times \vA$ is the magnetic field. Due to the use of the Weyl gauge Gau\ss{}' law is lost as equation of motion and has to be imposed as constraint to the wave functionals
\be
\label{2509-f49-x9}
\vec{\nabla} \vec{\Pi} | \psi \rangle = g \rho | \psi \rangle \, ,
\ee
where $\rho$ denotes the charge density of the matter fields. The operator $\vec{\nabla} \vec{\Pi} $ is the generator of time-independent gauge transformations (which are not fixed by the Weyl gauge $A^0 = 0$). In the absence of matter charges $(\rho = 0)$ Gau\ss{}' law implies gauge-invariance of the wave functional. Instead of working with explicitly gauge invariant wave functionals it is usually more convenient to explicitly resolve Gau\ss{}' law by fixing the gauge. For this purpose Coulomb gauge $\vec{\nabla} \vA = 0$ is particularly convenient since the remaining transversal components of the gauge field are the gauge-invariant degrees of freedom. In Coulomb gauge the transversal part of the momentum operator $\vec{\Pi}^\perp$ is still given by $\delta/\ii \delta \vA_\perp$ while for the longitudinal part from Gau\ss{}' law (\ref{2509-f49-x9}) follows
\be
\label{2518-f49-x10}
\vec{\Pi}^{||} | \psi \rangle = g \vec{\nabla} (- \Delta)^{- 1} \rho | \psi \rangle \, .
\ee
With this relation and with $\vec{\Pi}^2 = \vec{\Pi}^{\perp^2} + \vec{\Pi}^{||^2}$ one finds from eq.~(\ref{2495-f49-x7}) the gauge fixed Hamiltonian
\be
\label{2524-f49-x11}
H = \frac{1}{2} \int \lk \vec{\Pi}^{\perp^2} + \vB^2 \right) + \frac{g^2}{2} \int \rho (- \Delta)^{- 1} \rho \, .
\ee
Here the last term gives already the usual Coulomb interaction for a charge density $\rho$. Although the Hamiltonian approach yields the Coulomb law directly the latter can be also extracted from the spatial Wilson loop.

The exact vacuum wave functional which solves the Schr\"odinger equation with the Hamiltonian (\ref{2524-f49-x11}) for $\rho = 0$ is given by the Gaussian 
\be
\label{2538-f49-x12}
\psi (A) \equiv \langle A | \psi \rangle \sim \exp \left[- \int A \omega A \right] \, ,
\ee
where in momentum space $\omega (\vp) = | \vp|$ is the photon energy. It is clear that in the Hamiltonian approach, due to the use of the Weyl gauge $A_0 = 0$, only the spatial Wilson loop is accessible, for which one finds
\begin{align}
\label{1544-f49-x13}
\langle W (C) \rangle &= \langle \psi | W_C [A] | \psi \rangle \nonumber\\
&= \exp \left[ - \frac{g^2}{2} \oint\limits_C \dd x^i \oint\limits_C \dd y^j \, \cD_{ij} (\vx - \vy) \right] \, ,
\end{align}
where 
\begin{align}
\label{2550-f49-x14}
\cD_{ij} (\vx - \vy) &= \langle \psi | A_i (\vx) A_j (\vy) | \psi \rangle \nonumber\\
&= \frac{1}{2} t_{ij} (\vx) \omega^{- 1} (\vx, \vy) =  t_{ij} (\vx) \frac{1}{4 \pi^2 (\vx - \vy)^2}
\end{align}
is the static photon propagator with the 3-dimensional transversal projector $t_{ij} = \delta_{ij} - \nabla_i \nabla_j / \Delta$. This propagator also represents the equal-time limit of the time-dependent propagator in Landau gauge (\ref{2473-f45-x5})
\be
\label{2557-f49-x15}
\cD_{ij} (\vx - \vy) = \left. \cD_{\mu = i, \nu = j} (x - y) \right|_{x^0 = y^0} \, .
\ee
As in the time-dependent case, only the Kronecker part of the transversal projector contributes to the Wilson loop (\ref{1544-f49-x13}). The Coulomb law (\ref{2482-f45-x6})
can be also extracted from the spatial Wilson loop (\ref{1544-f49-x13}) by choosing a rectangular loop of sides $L$ and $R$ with $L \gg R$ and continuing $L$ to imaginary values $L = - \ii T$.

\section{The Wilson loop of QED in light-front quantization}\label{sectVI}

In section \ref{sectV} we have canonically quantized the photon sector of QED in the front-form and found that the vacuum state is given by the Fock-vacuum $| 0 \rangle$. We will now use this state to evaluate the spatial Wilson loop
\be
\label{2675-*3}
\langle W [A] \rangle = \langle 0 | W [A] | 0 \rangle \, .
\ee
In the common instant-form quantization the evaluation of the (spatial) Wilson loop in QED, reviewed in the previous section, is trivial since the vacuum wave functional is a Gaussian and the gauge fields at different space-time points commute so that the path ordering becomes irrelevant. Contrary to that in the light-front quantization the coordinate representation of the vacuum wave functional of QED is not at our disposal and the gauge field satisfies the non-trivial commutation relation (\ref{768-f13}).

As the calculation of the propagator given in section \ref{sectVC} taught us, we will have to start with a time-dependent loop and take the static limit $x^+ \to 0$ only at the 
very end of the calculation. Since we prefer to work with time-independent (stationary) wave functions the Wilson loop operator has to be taken in the Heisenberg picture. 

Consider a generic (time-dependent) Wilson loop $C$ in Minkowski space defined by eq.~(\ref{2254-41x1}), where $A_\mu (x)$ denotes the (time-dependent) gauge field in the Heisenberg picture (\ref{1393-x0}). Due to the chosen gauge (\ref{651-19}) only the projection $\vA_\perp$ of $A^\mu$ onto the transverse plane contributes to $W [A]$. Let ${x}^\mu (t)$ denote a parametrization of $C$ with $t \in [0, 1]$ and ${x}^\mu (1) = {x}^\mu (0)$. Then we have
\be
\label{1637-x2}
W [A] = T \exp \left[ \ii g \il^1_0 \dd t \,  \dot{{\vx}}_\perp (t) \cdot \vA_\perp \lk {x}^+ (t), {x}^- (t), \tilde{\vx}_\perp
(t) \right) \right] \, ,
\ee
where $T$ means (``time''-)ordering with respect to the  loop parameter $t$. Let us stress that the Wilson loop (\ref{1637-x2}) is different from the spatial Wilson loop
\be
\label{1644-x3}
\bar{W} [A] = P \exp \left[ \ii g \oint\limits_{\overline{C}} \dd \vx_\perp \cdot \vA_\perp (x^+ = 0, x^- = 0, \vx_\perp) \right] 
\ee
calculated from the projection $\overline{C}$ of $C$ onto the transversal subspace (1--2-plane). This is obvious when the Wilson loop obeys a perimeter law. 

The evaluation of the Wilson loop $\langle W [A] \rangle$ (\ref{1637-x2}) is non-trivial since in light-front quantization the gauge fields $\vA_\perp$ at different space-time points do not commute. However, the evaluation of $\langle W [A] \rangle$ is facilitated by the fact that the gauge potential (\ref{1412-4}) consists of two pieces
\be
\label{2698-4}
A^k (\vx) = \alpha_k (\vx) + \alpha^\dagger_k (\vx) 
\ee
with 
\be
\label{2703+-5}
\alpha_k (\vx) = \int \ddbar^3 p \frac{\Theta (p_-)}{\sqrt{2 p_-}} a_k (\vp) \efk^{- \ii p \cdot x} \, 
\ee
(where $p \cdot x$ is defined by eq.~(\ref{884-15-3}) with $p_+ = \vp^2_\perp / 2 p_-$) whose commutator  is a $c-$number. This is because $\alpha_k (\vx)$ and $\alpha^\dagger_k (\vx)$ are linear in the $a_k (\vp)$ and $a^\dagger_k (\vp)$, respectively, whose commutator (\ref{1500-f27}) is a $c$-number. The same is obviously true for the operators
\be
\label{2720-7}
\bar{\alpha} (t) = \dot{\vx} (t) \cdot \vec{\alpha} (\vx (t)) \, , \quad \quad \bar{\alpha}^\dagger (t) = \dot{\vx} (t) \cdot \vec{\alpha}^\dagger (\vx (t)) \, .
\ee
Then the following identity (which is an extension of Wick's theorem, see e.g.~\cite {R8}) holds
\begin{align}
\label{2709-6}
W [A] & \equiv T \exp \left[ \ii g \int \dd t \, (\bar{\alpha} (t) + \bar{\alpha}^\dagger (t)) \right] \nonumber\\
& = : \exp \left[\ii g \int \dd t \, (\bar{\alpha} (t) + \bar{\alpha}^\dagger (t)) \right] : \exp \Bigl[ - g^2 \int \dd t \int \dd t' \, \contraction{}{\bar{\alpha}^\dagger}{(t)}{\bar{\alpha}} {\bar{\alpha}^\dagger}{(t)}{\bar{\alpha}} (t') \Bigr]
\end{align}
Here $: \hk :$ denotes the normal ordered product with all operators $\bar{\alpha}^\dagger$ being placed left of the operators $\bar{\alpha}$. Furthermore `` $\!\!\!\contraction{}{\phantom{ab}}{\phantom{abc}}{} \phantom{abcde}$ '' denotes the contraction, which is defined as the difference between the ``time''-ordered and normal ordered product. 

Taking the vacuum expectation value of eq.~(\ref{2709-6}) we obtain for the Wilson loop
\be
\label{2727-8}
\langle W [A] \rangle = \exp \left[ - g^2 \int \dd t \int \dd \contraction{t'}{\bar{\alpha}^\dagger}{\bar{\alpha}}{(t')} {t'}{\bar{\alpha}^\dagger (t)}{\bar{\alpha}}{(t')} \right] = \exp \left[ - g^2 \int \dd t \int \dd t' \, \Theta (t' - t) [\bar{\alpha} (t'), \bar{\alpha}^\dagger (t) ] \right]  
\, .
\ee
The remaining commutator can be straightforwardly evaluated. Since it is a $c$-number we have
\be
\label{1831-f33a}
\left[ \alpha_k (x), \alpha^\dagger_l (y) \right] = \langle 0 | \left[ \alpha_k (x), \alpha^\dagger_l (y) \right] | 0 \rangle
\ee
and furthermore since $\alpha_k (x) | 0 \rangle = 0$ it follows
\be
\label{1836-f33a-1}
\left[ \alpha_k (x), \alpha^\dagger_l (y) \right] = \langle 0 | \alpha_k (x) \alpha^\dagger_l (y) | 0 \rangle = 
\langle 0 | A_k (x) A_l (y) | 0 \rangle
\ee
where the last two-point function was already evaluated in eq.~(\ref{1455-11}). Therefore we find for the Wilson loop (\ref{2727-8})
\be
\label{1728-3}
\langle W [A] \rangle = \exp \left[ - g^2 \int \dd t \int \dd t' \, \Theta (t' - t) \dot{x}^l (t') \dot{x}^k (t) 
\langle 0 | A^l (t') A^k (t) | 0 \rangle \right] \, .
\ee
Interchanging here the integration variables $t \longleftrightarrow t'$ and the summation indices $k \longleftrightarrow l$ in one half of the exponent we obtain the alternative expression
\be
\label{1741-y1}
\langle W [A] \rangle = \exp \left[ - \frac{g^2}{2} \int \dd t \int \dd t' \, \dot{x}^k (t) \dot{x}^l (t') \langle 0 | T_t \lk A^k (t) A^l (t') \right) | 0 \rangle \right] \, .
\ee
Choosing the parameter $t$ as the LF time $x^+$ the two-point function in the exponent becomes the photon propagator (\ref{1457-f26}) and we obtain
\be
\label{1866-f33b}
\langle W [A] \rangle = \exp \left[ - \frac{g^2}{2} \int \dd x^+ \int \dd y^+ \, \frac{\dd x^i}{\dd x^+} \frac{\dd y^j}{\dd y^+} \cD^{ij} (\vx (x^+), \vy (y^+)) \right] \, .
\ee
This result can be written in a parameter independent form 
\be
\label{1872-f33b-1}
\langle W [A] \rangle = \exp \left[ - \frac{g^2}{2} \oint_C \dd x^i \oint_C \dd y^j \, \cD^{ij} (x, y) \right] \, .
\ee
Choosing now a planar loop $C$ in the transverse plane (i.e.~$y^- = x^-$) eq.~(\ref{1872-f33b-1}) yields the same result as obtained in the canonical quantization in the instant form, see eq.~(\ref{1544-f49-x13}). Since also the photon propagators agree\footnote{Recall that only the Kronecker part $\delta_{kl}$ of the transverse projector $t_{kl} = \delta_{kl} - \partial_k \partial_l / \partial^2$ of the photon propagator $\cD_{kl}$ (\ref{2550-f49-x14}) contributes to the Wilson loop (\ref{1544-f49-x13}).} (c.f.~eqs.~(\ref{1477-x1}) and (\ref{2550-f49-x14})) we do find the same result for the Wilson loop in both approaches.

It is important to emphasize that the correct result for the Wilson loop is obtained in the LF quantization only by starting with temporal loops with non-trivial time-dependence and considering the static limit only at the very end of the calculation. The finite time interval serves as a regularization in the spirit of point splitting. Starting from the beginning with static, i.e.~spatial loops, leads to the ill-defined static propagator given in eq.~(\ref{1483-26}). 

\section{Conclusions}\label{sectVII}

In this paper we have investigated QED in the light-front quantization using Dirac's method for the quantization of constrained systems. Assuming the LF analogs of the Weyl gauge and of the axial gauge we have derived the gauge fixed LF Hamiltonian and evaluated from its vacuum state the static photon propagator in coordinate space as well as the spatial Wilson loop. We found that both quantities exist in the LF quantization only if the static limit $x^+ \to 0$ is taken in a specific way, i.e.~after all momentum integrals are carried out. Thereby the finite time argument $x^+$ serves as regulator in the spirit of point splitting.

Although in QED the vacuum wave functional is given by the Fock vacuum, the evaluation of vacuum expectation values turns out to be more involved than in the usual canonical quantization in the instant-form due to the non-commutativity of the photon field in light-front quantization. This calls into question the common lore that the LF vacuum is trivial. The problem concerning the static limit $x^+ \to 0$, which we have found here, should also manifest itself in the calculations of other time-independent quantities. 

\section*{Acknowledgment}
Useful discussions with M. Quandt are gratefully acknowledged. This work was supported in part by Deutsche Forschungsgemeinschaft under contract DFG-Re 856/10-1.

\begin{appendix}
\section{Relevant Poission brackets}\label{appC}

Below we list the Poisson brackets between the canonical variables and the constraints (\ref{567-7}), (\ref{604-12}) and (\ref{651-19})
\begin{align}
 \label{753-f13-x1}
\{ A^+ (x) , \varphi_1 (y) \} &= \{ A^+, \varphi_{2 k} \} = 0 \, , \nonumber\\
\{ A^+ (x) , \varphi_3 (z) \} &= - \partial^x_- \delta (x, z) \, , \nonumber\\
\{ A^+, \chi_1 \} &= 0 = \{ A^+, \chi_3 \} \, , \nonumber\\
\{ A^- (x) , \varphi_1 (y) \} &= \delta (x, y) \, , \quad \quad \{ A^-, \varphi_{2 k} \} = \{ A^-, \varphi_3 \}  = 0 \, , \nonumber\\
\{ A^-, \chi_1 \} &= 0 = \{ A^-, \chi_3 \} \, , \nonumber\\
\{ A^i (x), \varphi_1 \} &= 0 \, , \quad \quad \{ A^i (x), \varphi_{2 k} (y) \} = \delta^i_k \delta (x, y) \, , \nonumber\\
\{ A^i (x), \varphi_3 (y) \} &= \partial^x_i \delta (x, y) \, , \nonumber\\
\{ A^i (x), \chi_1 \} &= 0 = \{ A^i (x), \chi_3 \} \, , \nonumber\\
\{ \varphi_1, \Pi_+ \} &=  0 \, , \quad \quad \{ \varphi_3, \Pi_+ \} = 0 \, , \quad \quad \{ \chi_1, \Pi_+ \}  = 0 \,  , \nonumber\\
\{ \varphi_{2k} (x), \Pi_+ (y) \} &= - \partial^x_k \delta (x, y) \, , \nonumber\\
\{ \chi_3 (x), \Pi_+ (y) \} &= \partial^x_- \delta (x, y) \, , \nonumber\\
\{ \varphi_1, \Pi_- \} &= 0 \, , \quad \quad \{ \varphi_{2 k}, \Pi_- \} = 0 \, , \quad \quad \{ \varphi_3, \Pi_- \} = 0 \, , \nonumber\\
\{ \chi_1 (x), \Pi_- (y) \} &= \delta (x, y) \, , \quad \quad \{ \chi_3, \Pi_- \} = 0 \, , \nonumber\\
\{ \varphi_1, \Pi_i \} &=  0 \, , \quad \quad \{ \varphi_3, \Pi_i \} = 0 \, , \nonumber\\
\{ \varphi_{2 k} (x) , \Pi_i (y) \} &= - \partial^x_- \delta (x, y) \delta_{ki} \, , \nonumber\\
\{ \chi_1, \Pi_i \} &= 0 \, , \quad \quad \{ \chi_3, \Pi_i \} = 0 \, .
\end{align}

\end{appendix}

\bibliography{WilsonBib}

\end{document}